\documentclass{aa}

\input epsf 

\def\hc#1{\leavevmode\hbox to \hsize{\hss #1\hss}\leavevmode}

\begin{document}

\sloppy

\thesaurus{3 (9.03.2,13.07.2)}

\title{A search for gamma-ray emission from the Galactic plane in the
longitude range between $37^\circ$ and $43^\circ$}
\titlerunning{Gamma-ray emission from the Galactic plane}
\authorrunning{F. Aharonian et al.}

\author{F.A. Aharonian\inst{1},
A.G.~Akhperjanian\inst{7},
J.A.~Barrio\inst{2,3},
K.~Bernl\"ohr\inst{1},
O.~Bolz\inst{1},
H.~B\"orst\inst{5},
H.~Bojahr\inst{6},
J.L.~Contreras\inst{3},
J.~Cortina\inst{2},
S.~Denninghoff\inst{2}
V.~Fonseca\inst{3},
J.C.~Gonzalez\inst{3},
N.~G\"otting\inst{4},
G.~Heinzelmann\inst{4},
G.~Hermann\inst{1},
A.~Heusler\inst{1},
W.~Hofmann\inst{1},
D.~Horns\inst{4},
A.~Ibarra\inst{3},
C.~Iserlohe\inst{6},
I.~Jung\inst{1},
R.~Kankanyan\inst{1,7},
M.~Kestel\inst{2},
J.~Kettler\inst{1},
A.~Kohnle\inst{1},
A.~Konopelko\inst{1},
H.~Kornmeyer\inst{2},
D.~Kranich\inst{2},
H.~Krawczynski\inst{1,}$^\%$,
H.~Lampeitl\inst{1},
E.~Lorenz\inst{2},
F.~Lucarelli\inst{3},
N.~Magnussen\inst{6},
O.~Mang\inst{5},
H.~Meyer\inst{6},
R.~Mirzoyan\inst{2},
A.~Moralejo\inst{3},
L.~Padilla\inst{3},
M.~Panter\inst{1},
R.~Plaga\inst{2},
A.~Plyasheshnikov\inst{1,}$^\S$,
J.~Prahl\inst{4},
G.~P\"uhlhofer\inst{1},
W.~Rhode\inst{6},
A.~R\"ohring\inst{4},
G.P.~Rowell\inst{1},
V.~Sahakian\inst{7},
M.~Samorski\inst{5},
M.~Schilling\inst{5},
F.~Schr\"oder\inst{6},
M.~Siems\inst{5},
W.~Stamm\inst{5},
M.~Tluczykont\inst{4},
H.J.~V\"olk\inst{1},
C.A.~Wiedner\inst{1},
W.~Wittek\inst{2}}

\institute{Max Planck Institut f\"ur Kernphysik,
Postfach 103980, D-69029 Heidelberg, Germany \and
Max Planck Institut f\"ur Physik, F\"ohringer Ring
6, D-80805 M\"unchen, Germany \and
Universidad Complutense, Facultad de Ciencias
F\'{i}sicas, Ciudad Universitaria, E-28040 Madrid, Spain
\and
Universit\"at Hamburg, II. Institut f\"ur
Experimentalphysik, Luruper Chaussee 149,
D-22761 Hamburg, Germany \and
Universit\"at Kiel, Institut f\"ur Experimentelle und Angewandte Physik,
Leibnizstra{\ss}e 15-19, D-24118 Kiel, Germany\and
Universit\"at Wuppertal, Fachbereich Physik,
Gau{\ss}str.20, D-42097 Wuppertal, Germany \and
Yerevan Physics Institute, Alikhanian Br. 2, 375036 Yerevan,
Armenia\\
\hspace*{-4.04mm} $^\S\,$ On leave from
Altai State University, Dimitrov Street 66, 656099 Barnaul, Russia\\
\hspace*{-4.04mm} $^\%$ Now at Yale University, P.O. Box 208101, New Haven, CT 06520-8101, USA\\
}

\mail{Hubert Lampeitl, \\Tel.: (Germany) +6221 516 528,\\
email address: lampeitl@daniel.mpi-hd.mpg.de}

\offprints{Hubert Lampeitl}

\date{Received ; accepted }

\maketitle

\renewcommand\bottomfraction{0.90}

\begin{abstract}

Using the HEGRA system of imaging atmospheric Cherenkov telescopes,
a region of the Galactic plane ($-10^\circ < b < 5^\circ$, 
$38^\circ < l < 43^\circ$) was 
surveyed for TeV gamma-ray emission, both from point sources and 
of diffuse nature. The region covered includes 15 known pulsars, 6 known 
supernova remnants (SNR) and one unidentified EGRET source. 
No evidence for emission
from point sources was detected; upper limits are typically below
0.1 Crab units for the flux above 1~TeV. For the diffuse 
gamma-ray flux from the Galactic plane, an upper limit of
$6.1\cdot10^{-15}$ ph cm$^{-2}$ s$^{-1}$ sr$^{-1}$ MeV$^{-1}$
was derived under the assumption that the spatial distribution
measured by the EGRET instrument extends to the TeV regime.
This upper flux limit is a factor of about 1.5 larger than the
flux expected from the ensemble of gamma-ray unresolved Galactic 
cosmic ray sources.


\end{abstract}

\section{Introduction}

Systems of imaging atmospheric Cherenkov telescopes, such as the 
HEGRA stereoscopic telescope system (Daum et al. 1997, Konopelko et al. 1999a),
allow to reconstruct 
the directions of air showers over the full field of view -
with a radius of about 2$^\circ$ - and can therefore be used for
sky surveys (P\"uhlhofer et al. 1999). Here, we report on a 
survey of a rectangular
patch of the sky of roughly 80~deg$^2$, centered on the
Galactic plane at longitude $40^\circ$. The motivation for this
survey was twofold:
\begin{itemize}
\item Search for diffuse gamma-ray emission from the Galactic
plane
\item Search for gamma-ray point sources
\end{itemize}
Diffuse emission from the Galactic plane results from the
interactions of charged cosmic rays  
with interstellar gas confined to the plane or with photons. 
Diffuse emission 
in the energy range from tens of MeV to tens of GeV has been 
studied intensively by
SAS~2 (Fichtel et al. 1975; Hartmann et al. 1979),
COS~B (Mayer-Hasselwander et al. 1980, 1982)
and EGRET (Hunter et al. 1997). The basic features can be modeled assuming
$\pi^o$ decay as the dominant mechanism, with the gamma-ray emission
proportional to the product of the gas column density and 
the cosmic-ray density (see, e.g., Hunter et al. 1997, Fichtel et al. 1975,
Strong et al. 1988, Bloemen 1989, Bertsch et al. 1993).
The distribution of cosmic rays is assumed to follow the matter density 
with a characteristic correlation scale of 1.5 to 2 kpc
(Hunter et al. 1997). 
In addition to diffuse Galactic gamma rays, there is also a small 
extragalactic component, which should be fairly isotropic
(Fichtel et al. 1978; Sreekumar et al. 1998). Above 1 GeV, data show an
excess in gamma-ray flux over model predictions
(Hunter et al. 1997; see however Aharonian and Atoyan 2000). At these
energies, contributions from inverse Compton scattering
of electrons start to become relevant
(see, e.g., Porter \& Protheroe 1997). In response,
revised models speculate that the local measurements of 
the electron flux may not be representative for the entire
Galaxy (Porter \& Protheroe 1997; Pohl \& Esposito 1998). 
Electron propagation is limited by radiative
losses, and the local electron spectra are strongly influenced
by the history of sources in the local neighborhood 
(Aharonian, Atoyan \& V\"olk 1995; Pohl \& Esposito 1998).
In case that the solar system is in an ``electron void'', 
diffuse gamma-ray emission at high energies could be an
order of magnitude above predictions based on local electron
spectra (Pohl \& Esposito 1998).
Detailed models of the full spectrum of diffuse gamma-ray 
emission from the Galaxy (Moskalenko \& Strong 2000;
Strong, Moskalenko \& Reimer 1999, 2000) also favour a harder
electron spectrum. In the Galactic plane another `diffuse'
gamma-ray flux component arises from the hard energy spectrum
of those Galactic CRs that are still confined in the ensemble
of their unresolved sources. Assuming these to be SNRs, 
Berezhko \& V\"olk (2000) showed that their spatially averaged
contribution to the diffuse gamma-ray flux at 1~TeV should exceed
the model predictions of Hunter et al. (1997) by almost an order 
of magnitude.
It is therefore of significant interest to search
for the extension of the diffuse emission from the Galactic
plane at higher energies. Upper limits on diffuse gamma-ray
emission at TeV energies have been reported by
Reynolds et al. (1993), and LeBohec et al. (2000),
at a level of a few $10^{-3}$ of the cosmic-ray flux.
The most stringent limits at higher energies, above 100 TeV,
were given by Borione et al. (1997), and constrain the diffuse
flux to less than $3 \cdot 10^{-4}$ of the cosmic-ray flux.

The Galactic plane is also a region rich in potential 
gamma-ray point
sources. For obvious reasons, supernova remnants as well as 
pulsar driven nebulae cluster along the Galactic plane. Both types of objects
are almost certainly accelerators of cosmic rays and 
emitters of high-energy gamma radiation. Theoretical
models predict that typical gamma-ray fluxes from the majority
of these objects are below the detection threshold of the
current generation of instruments (see, e.g., Aharonian et al. 1997; Drury et al. 1994).
However, both the lack of knowledge of the individual source parameters as well as
approximations used in the modeling result in large
uncertainties in the predictions for individual objects, by an order of magnitude or more.
In addition to pulsars and supernova remnants, many
unidentified EGRET sources lie in the Galactic plane. Given the
density of source objects, a survey provides an efficient
way to search for gamma-ray emission.

The selection of the region to be surveyed was governed both 
by astrophysical and by practical considerations. Using the
EGRET observations of the diffuse gamma-ray flux (Hunter et al. 1997)
as a guideline,
a region within the inner Galaxy ($|l| < 40^\circ$)
and close to the Galactic equator
should show the strongest emission, corresponding to the gas 
column density. Similarly, the density of
supernova remnants (Green 1998)
and pulsars (Tayler 1993)
is highest in this region. 
However, from the location of the HEGRA telescope system,
at $28^\circ 45'$ N, the Galactic centre region can only be observed
at rather unfavourable zenith angles of 60$^\circ$ or more, 
and the observation time is rather limited. At such large
zenith angles, the energy threshold of the telescopes is
increased from 500 GeV for vertical showers to 5 TeV at 
60$^\circ$ (Konopelko et al. 1999b). 
Best sensitivity is obtained for observations at zenith angles
below 30$^\circ$, favouring regions at larger Galactic longitude.
Another potential problem for Cherenkov observations is 
the large and spatially varying background light from regions
of the Galactic plane, which may result in non-uniform 
sensitivity across the field of view of the Cherenkov telescopes.

As a result of these considerations, a rather dark region
at about 40$^\circ$ longitude was chosen for the survey, with an
extension in Galactic longitude of 5$^\circ$ and in latitude
of 15$^\circ$ (Fig.~\ref{fig_region}).
 The relatively large range in latitude was chosen
in order to cover the full range expected for diffuse gamma-ray emission, 
taking into account that the inverse Compton mechanisms may result 
in a wider latitude
distribution than observed in the EGRET data, and to include
additional background regions at large latitude. The observation
region includes part of the ``Sagittarius Arm'', one of the
spiral arms of the Galaxy. In addition, it hosts 15 known pulsars
(Table~\ref{tab_pulsar}),
6 supernova remnants (Table~\ref{tab_snr}), 
and the unidentified EGRET source 3EG J1903+0550
(Hartman et al. 1999).
There are no bright stars in the survey region, which
might seriously impact the observations; the brightest star is 
of $m_v=4.3^{mag}$, and there are 10 stars brighter than $6^{mag}$. (A $4.3^{mag}$ star
increases the DC current in a camera pixel from 0.8 to 10 $\mu$A, resulting 
in an increased noise of 1.5 ph.e. rms).
\section{The HEGRA IACT system}
\begin{figure}
\begin{center}
\mbox{
\epsfxsize9.0cm
\epsffile{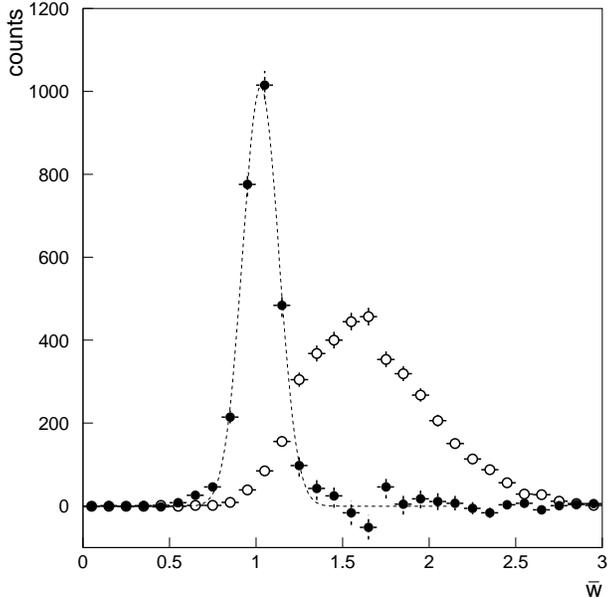}}
\caption{
Mean scaled width $\bar{w}$ of Cherenkov images in events where
at least four telescopes triggered, for $(\bullet)$ gamma-rays from the Crab
reference sample, and for $(\circ)$ cosmic-ray showers.}
\label{fig_mscw}
\end{center}
\end{figure}
The HEGRA stereoscopic system (Daum et al. 1997, Konopelko et al. 1999a)
of imaging atmospheric Cherenkov telescopes (IACTs)
is located on the Canary Island of La Palma, 
on the site of the Observatorio del Roque de los Muchachos,
at $28^\circ 45'$ N, $17^\circ 53'$ W, 2200 m a.s.l. 
The stereoscopic telescope system
comprises of five telescopes (CT2-CT6). One additional telescope
(CT1) is operated in stand-alone mode.
The system telescopes are arranged on the corners and in the centre of a
square with 100 m side length.
Each of them is equipped with tessalated 8.5~m$^2$ mirrors
of 5 m focal length, and a camera with 271 photomultiplier
pixels in the focal plane. The field of view of each camera
has 4.3$^\circ$ diameter; the pixel diameter corresponds to 0.25$^\circ$. 
The telescope system is triggered when in at least two
cameras two neighboring pixels show a signal above 6-8
photoelectrons. Signals from the cameras are recorded using 
a 120 MHz Flash-ADC system, which is read out after a trigger.
Details of the camera hardware and of the trigger system
are given in Hermann (1995) and Bulian et al. (1998). 
The pointing of the telescopes is known
better than 1 arcmin. (P\"uhlhofer et al. 1997). 
On the basis of
the stereoscopic analysis of Cherenkov images, shower
directions can be reconstructed with a precision of 0.1$^\circ$, 
and shower energies with a resolution of 20\% or better
(Daum et al. 1997; Aharonian et al. 1999a,b).
For vertical incidence of showers, the energy threshold is
500 GeV for gamma-rays, and increases to 0.9 TeV at 30$^\circ$ and
1.8 TeV at 45 $^\circ$ (Konopelko et al. 1999b). 
Cosmic ray showers are suppressed exploiting
the width of Cherenkov images. A ``mean scaled width''
$\bar{w}$ is defined by 
scaling the observed widths to the expected width for gamma-ray 
images, which depends on the intensity of the images
and the distance to the shower core, and averaging over
telescopes. Gamma-rays cause a peak in $\bar{w}$
at 1, with a Gaussian width of about 0.l to 0.12
(Fig.~\ref{fig_mscw}). Nucleonic
showers have larger $\bar{w}$ values, peaking around 1.7.
While more sophisticated identification schemes 
(e.g., Daum et al. 1997) can reach slightly better 
sensitivity, the default (and most stable) analysis schemes
are based on cuts in $\bar{w}$ ($\bar{w} < 1.1...1.3$), combined with an
angular cut relative to the source of about 0.1$^\circ$ 
\section{The data set}
The data used in this survey were taken during 37 days
in June, July and August 1999 with the complete 5-telescope system.

\begin{figure}
\begin{center}
\mbox{
\epsfxsize8.5cm
\epsffile{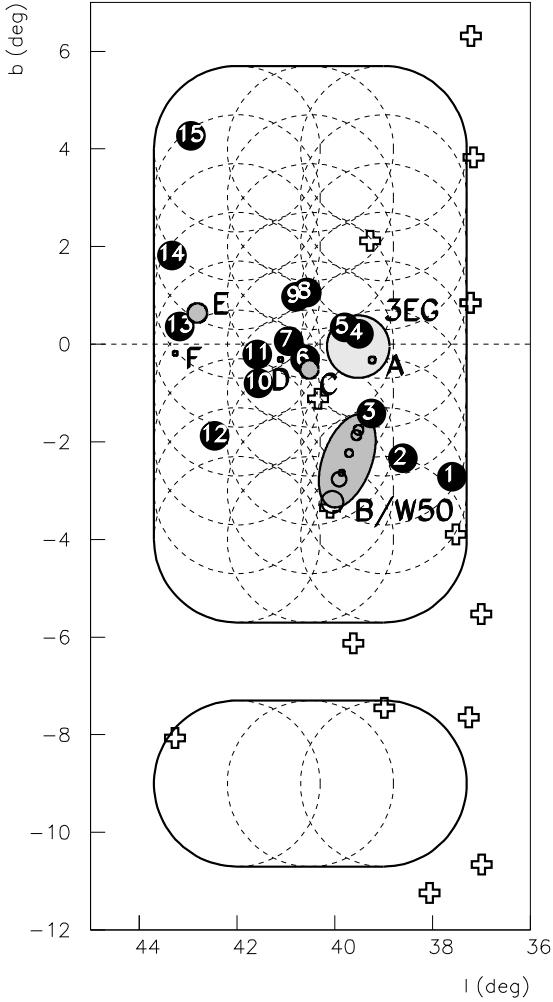}}
\caption[]{
Borders of the survey region (solid lines), and objects in this region.
The borders are shown for an effective field of view of 
1.7$^\circ$ radius, corresponding to the radius where the
detection efficiency is half of that compared
to the value on the optical axis of the telescopes. Individual scan
points are indicated as dashed circles.
Pulsars are shown as black circles, supernova remnants
as shaded regions, and stars brighter than $6^{mag}$ as crosses. The 
numbers of pulsars and SNR correspond to Tables~\ref{tab_pulsar}
and \ref{tab_snr}. The light gray area indicates the position of
the EGRET source 3EG J1903+0550 and the 1$\sigma$ error box.}
\end{center}
\label{fig_region}
\end{figure}

As illustrated in Fig. \ref{fig_region}, the total observation time of 
88~h was distributed between
three scans in Galactic latitude, centered at Galactic 
longitude of 39$^\circ$, 40.5$^\circ$ and 42$^\circ$, chosen
to guarantee an overlap of the effective fields of view
for the three scans (the scan positions given here and below
refer to the center of the field of view). 
Each scan was conducted in $1^\circ$
steps in latitude between $-4^\circ$ and $+4^\circ$, with an
additional control region at $-9^\circ$. The typical
time per scan point was about 15 min per day, and the scans were
repeated on several days. The points at latitude $0^\circ$
received twice the exposure, once covered 
going from $+4^\circ$ to $0^\circ$ and
once going from $-4^\circ$ to $0^\circ$.
A quality selection of the data sets was based primarily
on the average trigger rate of the telescope system.
A fraction of the data set suffered from extinction of
Cherenkov light due to Sahara dust in the atmosphere
(the so-called Calima, a well-known phenomenon at the
Canary Islands). The remaining data set is quite 
uniform in detection rate and not affected by Calima.
In total it encompasses 41.7 h,
18.6 h for the latitude scans at 39$^\circ$ longitude,
14.6 h at 40.5$^\circ$ longitude and 8.5 h at 42$^\circ$
longitude. The data cover zenith angles between 
$21^\circ$ and $35^\circ$; the zenith angle of a run and the
Galactic latitude are somewhat correlated, with data 
at $4^\circ$ latitude covering zenith angles between 20$^\circ$ and
$28^\circ$, compared to 28$^\circ$ to 35 $^\circ$ for the runs at
$-9^\circ$ latitude. In total, $1.4\cdot10^6$ Cherenkov events are used.

The analysis of Cherenkov images could potentially suffer from
variations in the sky brightness over the scan region. Since 
the readout electronics of the telescopes is AC coupled,
a star illuminating a pixel will not cause baseline shifts,
but it will still result in increased noise in that pixel.
As a 
measure of the influence of night sky background light, the 
baseline noise of the photomultiplier signal was determined.
The RMS of the baseline noise of the individual PMTs is 
quite homogeneous over the region and amounts in average to about 
1 ADC count. The brightest stars in the region increase the RMS 
to 1.5 ADC counts; this is uncritical for further image analysis.
For reference and comparison, observations of the Crab Nebula
in the winters 1997/1998, 1998/1999 and 1999/2000 were used, and were
subject to identical selection criteria. In total, 41.7 h of 
observations at zenith angles between 18$^\circ$ and 32$^\circ$
were used, with a roughly uniform coverage of the zenith angle
range.

The last  telescope integrated into the system, CT2, one of the corner 
telescopes, was only used for part of the Crab reference data sets.
Therefore, all analyses
discussed in the following were performed both using only
the four telescopes, as well as the full set of five telescopes.
Because of the good agreement between the simulations and the
measured rate with the complete system (see below), 
the five-telescope limits are
quoted as the final results.

\section{Search for gamma-ray point sources}
\begin{table}[tb]
\begin{center}
\caption{Known pulsars in the survey region. Columns include the 
Galactic longitude $l$ and latitude $b$, the period $P$
and the period derivative $\dot{P}$. The dispersion measure (DM)
can be used to estimate the distance to the pulsar, assuming
a uniform density of thermal electrons in the Galaxy, in the order of
0.1-0.01 cm$^{-3}$ and dividing the DM by this number. 
Data taken from Tayler (1993).}
\vspace{0.3cm}
\begin{tabular}{|r|l|c|c|c|c|c|}
\hline
  &Name PSR& $l [^\circ] $ & $b [^\circ]$ & $P$ [s]   & $\dot{P}$     & $DM$  \\
  &J(2000) &          &          &        &[$10^{-15}$]&[cm$^{-3}$pc]\\ \hline
1 & 1909+0254        & 37.6     & -2.7     & 0.99 & 5.5287         & 172.1   \\
2 & 1910+0358        & 38.6     & -2.3     & 2.33 & 4.53           & 78.8    \\
3 & 1908+04          & 39.2     & -1.4     & 0.29 & -              & 217     \\
4 & 1902+0556        & 39.5     & +0.2     & 0.75 & 12.896         & 179.7   \\
5 & 1902+06          & 39.8     & +0.3     & 0.67 & -              & 530     \\
6 & 1906+0641        & 40.6     & -0.3     & 0.27 & 2.1352         & 473     \\
7 & 1905+0709        & 40.9     & +0.1     & 0.65 & 4.92           & 269     \\
8 & 1901+0716        & 40.6     & +1.1     & 0.64 & 2.40           & 253     \\
9 & 1902+07          & 40.8     & +1.0     & 0.49 & -              & 90      \\
10& 1910+07          & 41.6     & -0.8     & 2.71 & -              & 115     \\
11& 1908+07          & 41.6     & -0.2     & 0.21 & -              & 10      \\
12& 1915+07          & 42.5     & -1.9     & 1.54 & -              & 50      \\
13& 1908+0916        & 43.2     & +0.4     & 0.83 & 0.098          & 250     \\
14& 1904+10          & 43.3     & +1.8     & 1.86 & -              & 140     \\
15& 1854+10          & 42.9     & +4.3     & 0.57 &                & 250     \\
\hline
\end{tabular}
\end{center}
\label{tab_pulsar}
\end{table}
\begin{table}[h]
\begin{center}
\caption{Known supernova remnants in the survey region (Green 1998). 
Most are shell-type
SNR. For W50 individual points known as X-ray emitters are given 
(see e.g. Safi-Harb et al. 1997). For the unidentified EGRET 
source the location and $1\sigma$ error box are given 
(Hartmann et al. 1999).}
\vspace{0.3cm}
\begin{tabular}{|l|l l|c|c|c|}
\hline
&Name & & $l [^\circ]$ & $b [^\circ]$ & diameter $[^\circ]$ \\
\hline
\hline
A & 3C396     &        & 39.2   & -0.3   & $\sim$ 0.13         \\
\hline
B & W50       &        & 39.7   & -2.0   & $\sim$ 2 $\times$ 1 \\
  &           & SS-433 & 39.71  & -2.25       &                \\
  &           & e1     & 39.75  & -2.66       & $\sim$0.12     \\
  &           & e2     & 39.91  & -2.80       & $\sim$0.3   \\
  &           & e3     & 40.09  & -3.21       & $\sim$0.45  \\
  &           & w2     & 39.51  & -1.76       &                \\
\hline
C & -         &   & 40.5   & -0.5   & $\sim$ 0.4          \\
D & 3C397     &   & 41.1   & -0.3   & $<$ 0.1             \\
E & -         &   & 42.8   & +0.6   & 0.4                 \\
F & W49B      &   & 43.3   & -0.2   & $<$ 0.1             \\
\hline
\hline
3EG           &\multicolumn{2}{l|}{3EG J1903+0550}& 39.52 & -0.05 & $1\sigma$ 0.64 \\
\hline
\end{tabular}
\label{tab_snr}
\end{center}
\end{table}
\begin{figure}
\begin{center}
\mbox{
\epsfxsize9.0cm
\epsffile{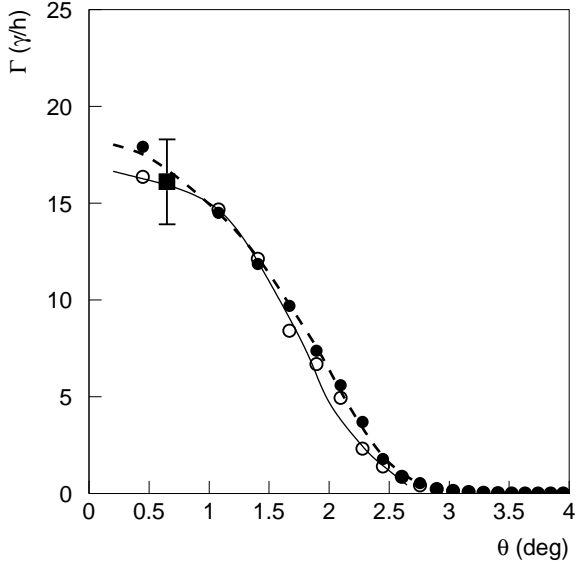}}
\caption{Detection rates for a point source with the flux and spectrum
of Crab Nebula, as a function of the angle of incidence $\Theta$ of the
photons relative to the telescope axis. The simulations were
performed for a five telescope system, and are shown for events
where all telescopes trigger. The rates include a cut on the
mean scaled width $< 1.1$. Also shown is the rate measured
for the Crab Nebula, positioned 0.5$^\circ$ off-axis. Full symbols:
20$^\circ$ zenith angle, open symbols: 30$^\circ$}
\label{fig_rates}
\end{center}
\end{figure}
\begin{figure}[t]
\begin{center}
\mbox{
\epsfxsize8.5cm
\epsffile{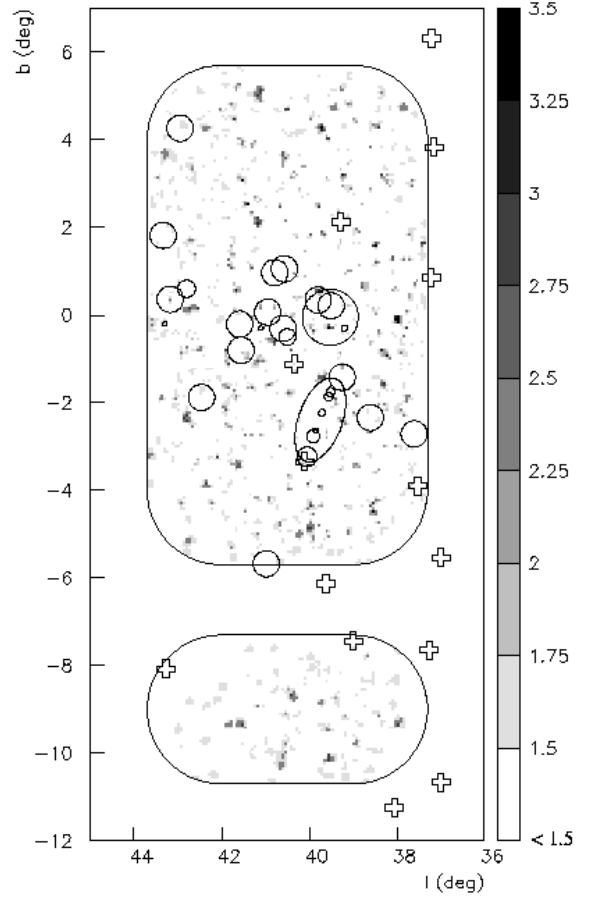}}
\caption{Map of the significances. Only significances above
1.5 sigma are shown. Superimposed are the locations of
potential sources 
and the borderline of the observed region 
(see also Fig.~\ref{fig_region}). 
}
\label{fig_sigmap}
\end{center}
\end{figure}
The cuts on the telescope images and the shower reconstruction
follow earlier work. In particular, only images with at least
40 photoelectrons are accepted and the centroid of the 
image has to be within  1.7$^\circ$ from the camera centre, in
order to exclude truncated images. Showers with reconstructed
cores up to 300 m from the center telescope are accepted.

The performance of the telescope system for off-axis gamma-rays
was investigated in detail using Monte Carlo simulations,
for the zenith angle range in question. Fig.~\ref{fig_rates} shows, 
as a function of shower inclination relative to the telescope
axis, the rate at which showers are triggered and reconstructed
with all five telescopes for a Crab-like source.
The  
telescope system shows a FWHM field-of-view of about 3.5$^\circ$.
Also shown is the actual Crab rate derived from 4.1~h of
observations. During pointed
observations with the HEGRA IACT system, the source is usually
positioned $0.5^\circ$ off-center. 
The angular resolution is, within 20\%, independent of
the inclination of the shower axis and varies between
about 0.11$^\circ$ for events triggered by at least two of the
five telescopes, to 0.07$^\circ$ for five-telescope events.  
The angular resolution is  measured by the Gaussian width 
of a 1-dimensional projected angular distribution,
corresponding to the 40\% containment radius 
\footnote{Non Gaussian tails depend on the number of telescopes 
used for reconstruction and are well below 10\% for the 
5-telescope events..} 
\renewcommand\textfraction{.05}
\begin{figure}[t]
\begin{center}
\mbox{
\epsfxsize8.8cm
\epsffile{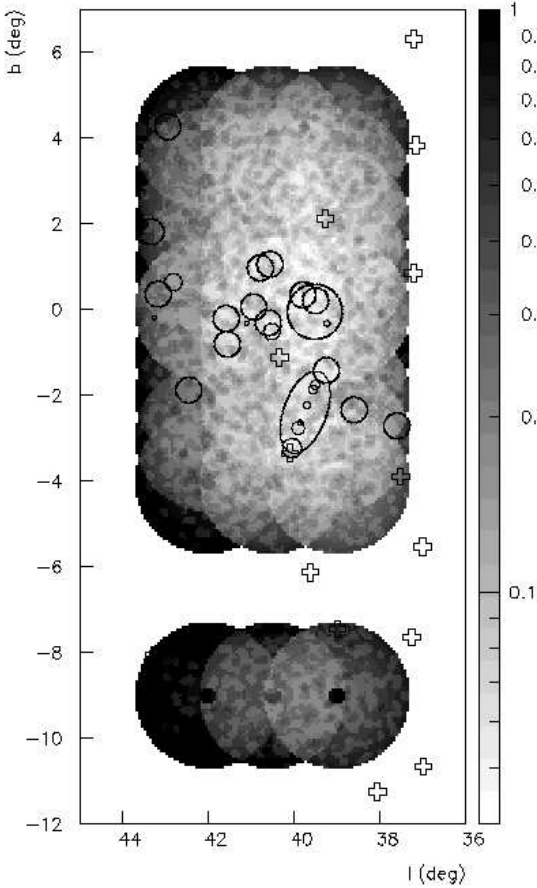}}
\caption{ 
Upper limits for the gamma-ray flux above 1 TeV, 
in units of the Crab flux. Upper limits above the
flux of the Crab are shown in black.
}
\label{fig_uplmap}
\end{center}
\end{figure}
\begin{table}[htb]
\begin{center}
\caption{Upper limits for specific objects. Columns show the 
observation time used in the analysis, the number of 
ON and OFF-source events (for a three times larger OFF
region), the significance of detection and the 99\%
upper limit in units of the Crab flux. 
Objects are treated as point sources (type P) or extended sources 
(type E1: $\vartheta = 0.2$ deg, type E2:$\vartheta = 0.25$ deg).}
\vspace{0.3cm}
\begin{tabular}{|l|c|c|c|c|c|c|}
\hline
Object     & Type  &T [h]  & ON   & OFF  & $\sigma$ & $\phi^{99\%}$ \\ \hline
J1909+0254 & P     &3.87  &    0 &    2 &  -1.07 & 0.150 \\ 
J1910+0358 & P     &6.93  &    1 &    6 &  -0.70 & 0.086 \\ 
J1908+04   & P     &10.75 &    0 &   10 &  -2.40 & 0.046 \\ 
J1902+0556 & P     &11.94 &    2 &    4 &  0.46  & 0.059 \\ 
J1902+06   & P     &11.94 &    3 &   11 &  -0.31 & 0.060 \\ 
J1906+0641 & P     &12.46 &    1 &    5 &  -0.49 & 0.051 \\ 
J1905+0709 & P     &8.78  &    2 &    7 &  -0.19 & 0.074 \\ 
J1901+0716 & P     &7.06  &    2 &    5 &  0.21  & 0.106 \\ 
J1902+07   & P     &8.97  &    2 &    3 &  0.74  & 0.083 \\ 
J1910+07   & P     &8.40  &    2 &    6 &  0.00  & 0.087 \\ 
J1908+07   & P     &8.78  &    1 &    6 &  -0.70 & 0.068 \\ 
J1915+07   & P     &1.99  &    0 &    2 &  -1.07 & 0.214 \\ 
J1908+0916 & P     &2.70  &    2 &    0 &  2.35  & 0.321 \\ 
J1904+10   & P     &1.55  &    0 &    0 &  0.00  & 0.533 \\ 
J1854+10   & P     &1.80  &    0 &    1 &  -0.76 & 0.285 \\ 
\hline
SNR-3C396  & P     &12.31 &    4 &    6 &   1.04 & 0.082 \\
SNR-G40.5  & E2     &9.62  &   30 &  108 &  -0.90 & 0.062 \\
SNR-3C397  & P     &8.78  &    3 &    4 &   1.03 & 0.096 \\
SNR-G42.8  & E2    &2.49  &   12 &   24 &   1.12 & 0.263 \\
SNR-W49B   & P     &2.16  &    1 &    0 &   1.67 & 0.337 \\

\hline
SS433       & P &    9.02 &    1 &    6 &-0.70 & 0.063 \\
SS433-e1    & P &   10.00 &    2 &    2 & 1.07 & 0.080 \\
SS433-e2    & E1 &   10.00 &   20 &   66 &-0.38 & 0.110 \\
SS433-e3    & E2 &    7.02 &   19 &   60 &-0.20 & 0.091\\
SS433-w2    & P &    9.02 &    2 &    4 & 0.46 & 0.081 \\                              
\hline
3EG         & E2 &12.83 &   39 &  124 &  -0.32 & 0.073\\      
\hline
\end{tabular}
\label{tab_upperlimit}
\end{center}
\end{table}
Since events with five triggered telescopes have both a better
angular resolution and an improved hadron rejection compared
to events with fewer telescopes, they contribute the bulk of 
the significance in the detection of sources. In fact, 
indiscriminate combination of 2, 3, 4 and 5-telescope events may 
deteriorate the sensitivity compared to 5-telescope events alone.
For these reasons, the current analysis is based exclusively on
events with five triggered telescopes for point sources
and on four and five triggered telescopes for extended sources.

For a given point source candidate, events reconstructed within 0.11$^\circ$
of the source direction were counted, after applying a cut
on $\bar{w}$ at 1.1 to reject cosmic-ray events. 
In case of extended objects, such as SNR, the angular cut
was increased to 0.2$^\circ$ or 0.25$^\circ$ depending on the source size, 
to ensure that the whole source is contained in the search region
and in addition the 4-telescope events were used, since angular
resolution is no longer critical. 
To estimate backgrounds, three
regions of the same size as the source region were used, 
rotated by 90$^\circ$,
180$^\circ$ and 270$^\circ$ around the telescope axis, relative
to the source. This background estimate can only be applied
for sources more than $0.12^\circ$ away from the telescope
axis (otherwise the source region and the background regions
would overlap). On-axis source regions are therefore excluded.
A source region which is on-axis for one scan point will, of
course, be off-axis for the neighbouring scan points, hence the
net loss is small. The significance for a detection is then
calculated according to Li \& Ma (1983), with $\alpha=1/3$ corresponding
to the three background regions per source region. \\
\renewcommand\textfraction{0.00001}
\renewcommand\floatpagefraction{1.1}
In a first search for gamma-ray sources at arbitrary locations,
significances were calculated for a two-dimensional grid with 
0.0625$^\circ$ step size, covering the survey region. Note that
the step size is smaller than the angular resolution to ensure
that no sources are missed. This implies, however, that 
significances quoted for adjacent points are highly correlated.
The highest 
significance observed is 4.1$\sigma$, compatible with the expected
distribution in the absence of sources. The mean significance is
0.006, the rms width of the distribution 1.027, which shows
that the technique used for background estimates is
sufficiently accurate.
A significance map is shown in Fig.~\ref{fig_sigmap}.
The points with the highest significances do not correspond to
locations of
any known potential source as listed 
in Tables~\ref{tab_pulsar} and \ref{tab_snr}. 
Upper limits are calculated according to the procedure
given in O. Helene (1983) and dividing this number by the 
expected number of events for a Crab like source seen at the same angle 
$\theta$ in the camera (see Fig. \ref{fig_rates}).
To predict the expected event rate  
the HEGRA Monte Carlo code (Konopelko, 1999a) were used to calculate
the acceptance over the FoV.
In the simulations the Crab spectrum is taken to be 
$d\Phi/dE = 2.7(\pm 0.2 \pm 0.8)\cdot 10^{-11} E^{-2.59 (\pm 0.06 \pm 0.1)}$
ph cm$^{-2}$ s$^{-1}$ Tev$^{-1}$ (Konopelko, 1998).
An additional scaling of the upper limit by 28\% accounts for losses 
caused by 
the tight angular cut for point sources.
The resulting upper limits
are shown in Fig.~\ref{fig_uplmap}. 
For the well-sampled
points of the survey the limits are below 10\% of the Crab 
flux. The limits given assume a slope of the energy spectrum similar to
the Crab Nebula. However, since limits are given for an
energy which corresponds to the peak differential detection rate,
the variation of the limits with the assumed spectral index 
is small (Aharonian et al. 1995). For a variation of $\pm0.5$ in the 
spectral index, the 
limits typically change by 6\%.
\\
Specific flux limits were derived for the potential source
candidates, namely the 15 pulsars, the 6 supernova remnants,
and the EGRET source. In case of the fairly extended SNR
W50, hot spots known from X-ray measurements were treated as
point sources. The so called eastern lobe (e3, see Tab. \ref{tab_snr})
and the associated knot e2 were treated as extended sources.
W50 has been observed previously by HEGRA and a more detailed analysis 
of the previous data set and a discussion of results on W50 will be 
given in a separate paper (Aharonian et al. 2001).\\
Table~\ref{tab_upperlimit}
lists the upper limits obtained for all objects.

\section{Search for diffuse gamma-ray emission from the Galactic plane}

\begin{figure}
\begin{center}
\mbox{
\epsfxsize9.0cm
\epsffile{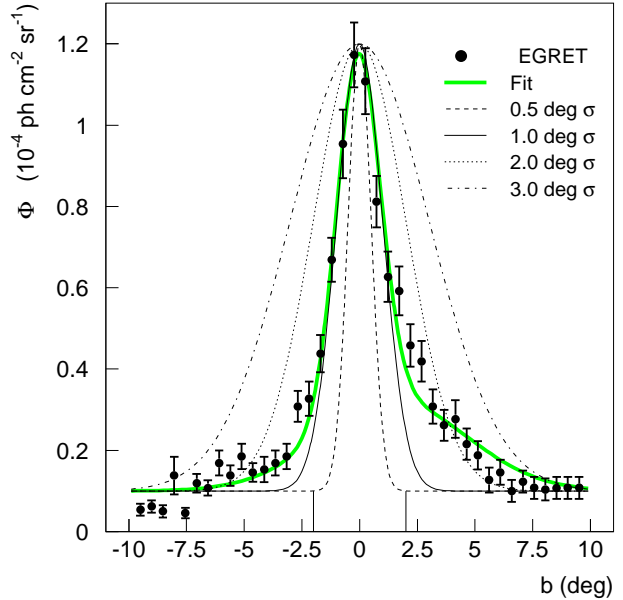}}
\caption{Latitude dependence of the diffuse emission measured by EGRET
for photon energies above 1 GeV,
in the range of Galactic longitude 30$^\circ$ to 50$^\circ$,
together with a fit by a sum of two Gaussians with widths and centres of 
$0.95^\circ$, $3.16^\circ$ and $-0.05^\circ$, $1.5^\circ$, 
respectively, and an intensity ratio of 4:1 in amplitude. A value of
0.1 was taken for the constant background. 
Data points taken from Hunter et al. 1997.
In addition, Gaussian profiles with a $\sigma$ of $0.5^\circ$ ,$1.0^\circ$,
$2.0^\circ$ and $3.0^\circ$ are shown.}
\label{fig_difflat}
\end{center}
\end{figure}

\begin{figure}
\begin{center}
\mbox{
\epsfxsize9.0cm
\epsffile{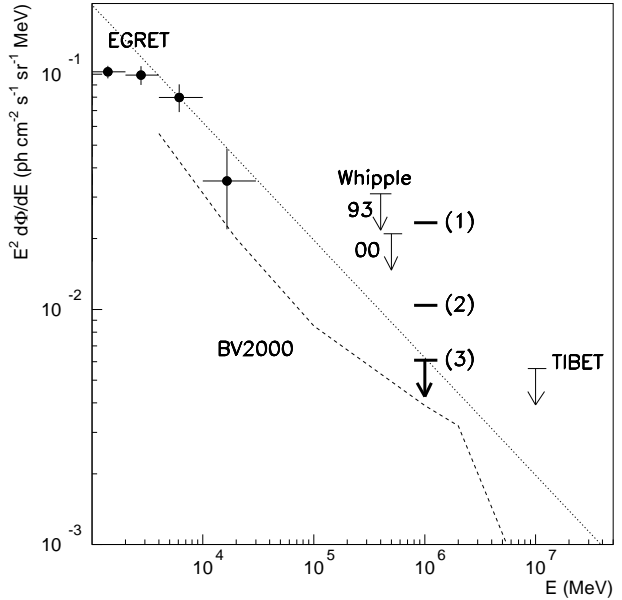}}
\caption{
Upper limits for the diffuse gamma-ray flux derived                          
by this experiment. Assuming that all detected events are gamma-rays (1), 
using a independent data set for background subtraction (2),
and using $|b|>2^\circ$ as background data and assuming the 
spatial distribution measured by the ERGRET instrument
($38^\circ < l < 43^\circ, |b| < 2 ^\circ$) (3). For detailed explanations
see text.
Also shown is the EGRET flux for $35^\circ < l < 45^\circ,
|b| < 2^\circ$, the Whipple upper limits for                            
$38.5 < l < 41.5^\circ, |b| < 2^\circ$ (Reynolds et al. 1993, LeBohec 2000)
and the Tibet upper limit (Amenomori, 1997). 
The dotted line indicates an extrapolation of the EGRET data 
with an index of 2.5. The dashed line indicates the             
scaled 'leaky box' model prediction by Berezhko \& V\"olk, 2000 (BV2000) 
for the spatially averaged gamma-ray emission from the Galactic plane
by those Galactic CRs that are still confined in their unresolved SNR sources,
scaled to $|b|<2^\circ$.}
\label{fig_results} 
\end{center}       
\end{figure}

Compared to the search for point sources, 
the search for diffuse gamma-ray emission from the Galactic plane
is complicated by the extended structure of the emission region.
The structure may be extended in latitude beyond the field of view.
Fig.~\ref{fig_difflat} illustrates the profile in 
Galactic latitude as measured by EGRET in the longitude range
30$^\circ$ to 50$^\circ$, at energies above 1 GeV. The EGRET
latitude profile can be represented as a sum of two Gaussians
and an constant background value. Since the angular resolution
of the EGRET instrument above 1 GeV is narrower than the latitude
extend of the diffuse emission the latitude  dependence in 
Fig.~\ref{fig_difflat} indicates real structure.
Because of the possible change of the primary emission mechanism
between the GeV and TeV range, from $\pi^0$ decay to inverse 
Compton scattering (Porter \& Protheroe, 1997; Pohl \& Esposito, 1998), 
and because of the completely different
parent and target populations in the two cases, the 
latitude dependence may differ at TeV energies. At GeV
energies, the latitude distribution of gamma-ray emissivity
is characterized by the scale height of interstellar gas,
around 100 pc for atomic hydrogen (Lockman 1984) and somewhat less for
molecular hydrogen (Dame 1987). In contrast, inverse-Compton
interactions with the uniform microwave background photons are governed
by the scale height of the electron component of cosmic rays,
which might be characterized by the kpc scale describing the coupling
of cosmic rays to matter. The scale height of far-infrared
target photons, on the other hand, was given by Cox, Kr\"ugel
and Metzger (1986) as ~100 pc, similar to the scale height of
the gas. Porter and Protheroe (1997)
find, from numerical simulations of electron propagation, 
a scale height of more than 500 pc for the inverse-Compton
emissivity at 1 TeV.\\
In the following, we will discuss three techniques to search
for diffuse gamma-ray emission. The techniques differ in the
degree of assumptions they make concerning the latitude dependency
of the diffuse emission, and also in their sensitivity to 
systematic variations in the performance and characteristics
of the telescopes.\\
\begin{figure}
\begin{center}
\mbox{
\epsfxsize9.0cm
\epsffile{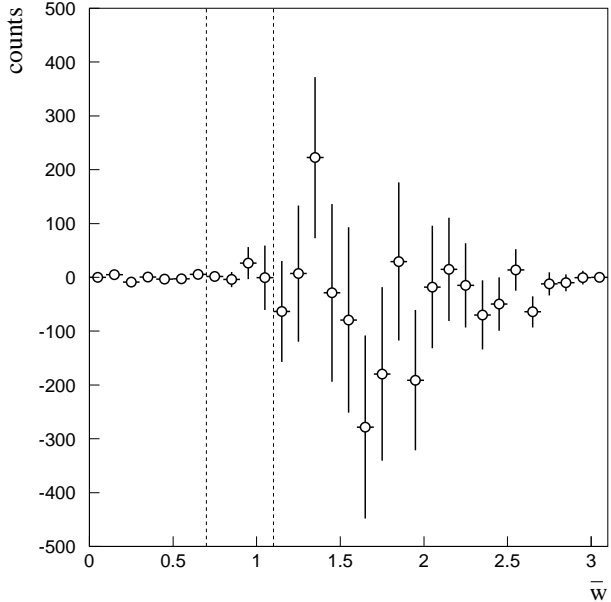}}
\caption{Difference between the distributions in mean scaled 
width for the on-region ($|b| < 2^\circ$) and for the off-region
($|b| > 2^\circ$). The dashed lines indicate the expected gamma-ray
region.}
\label{fig_mscwdis}
\end{center}
\end{figure}
The most robust and model-independent -- but also least sensitive --
technique to derive limits on the diffuse flux
simply selects events according to their shapes
as gamma-ray candidates. A cut on $\bar{w}$ less than 
1.0 keeps about 1/2 of the gamma-rays, but rejects cosmic rays very 
efficiently (see Fig. \ref{fig_mscw}). 
In order to achieve the best separation between gamma-ray
images and cosmic-ray images, only five-telescope events were used.
Such events with small $\bar{w}$ include genuine
gamma-rays, electron showers from the diffuse cosmic-ray electron flux,
and the tail of the distribution of cosmic-ray nuclear showers. 
Assigning all 428 events after cut as diffuse gamma-rays a 99\% upper 
limit on the diffuse gamma-ray flux at 1 TeV of 
23.4 $\cdot 10^{-15}$ ph cm$^{-2}$ s$^{-1}$ sr$^{-1}$ MeV$^{-1}$
results, for $|b| < 5^\circ$. As mentioned in the discussion of the 
point source limits, the measurement determines the integral flux above 
the energy threshold of the telescope system, rather than
directly determining the differential flux. Therefore the result
depends in principle on the assumed spectral index. However,
since flux values are quoted at energies corresponding roughly to the peak
detection rates, the limits vary only very little with the spectral
index. We note that the diffuse
electron flux (see, e.g., the compilation by Barwick et al. 1998) 
should contribute about 1/3 of the number
of gamma-ray candidates; the limit can of course also be viewed
as a limit on the electron flux, since the selection cuts are
equally efficient for gamma-ray induced showers and electron-induced
showers.\\
The limit obtained by this technique can be improved by subtracting,
on a statistical basis, the number of gamma-ray candidate events 
resulting from cosmic-ray electrons or protons. Since it is virtually
impossible to verify that simulations properly account for the tail
towards very small $\bar{w}(<1.1)$ of the distribution of
proton showers, such a subtraction has to be performed using an
experimental background region, sufficiently far away from the Galactic plane 
such that diffuse gamma-ray emission from the plane is most probably 
negligible. Such a background data sample will only contain the 
isotropic electron 
and nucleon flux. In order to minimize instrumental effects, 
the background sample
should be taken at the same time, and at identical zenith angles.
Data sets can be normalized to each other using the rates of events with large 
$\bar{w}$ ($> 1.4$), well outside the gamma-ray region.
Unfortunately, the availability of suitable background data samples
with the same telescope configuration 
is limited to 4.1 h of data,
and the statistical error of the background data set limits the sensitivity.
After a $\bar{w}<1.1$ cut 1928 events survived compared 
to 141 events in the referenc sample. The scaling factor between the two datasets is 
determined to 13.3 and the MC simulations predict 555 events for a Crab like
source smeared out over a FoV of $1.5^\circ$.
After subtraction of isotropic components, we find a 99\% limit on the 
diffuse flux in the region $|b| < 5^\circ$ of
10.4 $\cdot 10^{-15}$ ph cm$^{-2}$ s$^{-1}$ sr$^{-1}$ MeV$^{-1}$ at 1 TeV.\\
The final, and most sensitive analysis makes the assumption that
diffuse gamma-ray emission from the Galactic plane is limited
to the central parts of the scan region, and uses the outer
parts of the scan region to estimate backgrounds. 
The range $|b| < 2^\circ$ was considered the 
signal region, the range $|b| > 2^\circ$ the background region.
This cut is close to optimal for emission profiles with an
rms width between 1$^\circ$ and 2$^\circ$ and results in
a good balance in observation time 
($\alpha \approx 1$) between the ON and OFF regions. 
To ensure optimum quality of the events, only four- and five-telescope
events were used, and the field of view was restricted to 
1.5$^\circ$ from the optical axis. A cut at 1.1 on the
$\bar{w}$ was applied to reject cosmic-ray background
\footnote{The cuts on $\bar{w}$ differ between this
and the first analysis; here, the goal is to optimize the significance
of a weak signal $S$, e.g., the ratio $S/\sqrt{B}$, whereas analysis of
the first type, where all backgrounds $B$ are counted as potential signal,
need to optimize $S/B$.}.
To account for a possible zenith-angle dependence of 
background rates, data were grouped into four different
ranges in zenith angle, 20$^\circ$-24$^\circ$, 24$^\circ$-28$^\circ$, 
28$^\circ$-32$^\circ$
and 32$^\circ$-36$^\circ$. Also, the analysis was carried out
separately for each scan band. For each range in zenith angle and
each scan band, the expected number of events in the signal 
region was calculated based on the number of events observed
in the corresponding areas of the camera for the background 
region. The expected and observed numbers of events were then
added up for all zenith angles and scan bands.
With a total number of 2387 gamma-ray events in the signal region,
compared to 2353 expected events, there is no significant 
excess. 
As a cross check, Fig.~\ref{fig_mscwdis}
shows the background-subtracted distribution in $\bar{w}$
for events in the $|b| < 2^\circ$ region. There is no indication
of a significant excess
in the gamma-ray region around a $\bar{w}$ of 1 indicated by the dashed lines. 
Also for larger values of $\bar{w}$, there is no significant
excess or deficit, showing that background subtraction works properly.\\
In order to translate the limit in the number of events into 
a flux limit, one now has to make assumptions concerning the 
distribution in Galactic latitude of the diffuse radiation,
since a spill-over of diffuse gamma-rays into the background
region $|b| > 2^\circ$ will effectively reduce the signal.
For a profile with a width less or comparable to the
EGRET profile -- about $1^\circ$ rms -- a correction of 12\% is applied
and one finds a limit 
$6.1\cdot10^{-15}$ ph cm$^{-2}$ s$^{-1}$ sr$^{-1}$ MeV$^{-1}$ for the 
diffuse gamma-ray flux at 1 TeV, 
averaged over the $|b| < 2^\circ$ region.
The limit
refers to an assumed spectral index of -2.6, and changes 
by +13\% for an index of -2, and by -5\% for an index of -3.
For wider distributions of $2^\circ$ and $3^\circ$ rms, the
limit changes to $8.2 \cdot10^{-15}$ ph cm$^{-2}$ s$^{-1}$ sr$^{-1}$ MeV$^{-1}$
and $12.8\cdot10^{-15}$ ph cm$^{-2}$ s$^{-1}$ sr$^{-1}$ MeV$^{-1}$, 
respectively.
Fig.~\ref{fig_results} compares the upper limit with the
extrapolation of the EGRET flux; also included are the 
limits by the Whipple group (Reynolds et al. 1993; LeBohec et al. 2000)
and of the Tibet array (Amenomori, 1997).
Connecting the EGRET points with the HEGRA upper limit
(and ignoring the highest-energy EGRET point with its
large errors) one finds a lower limit of 2.5 on the spectral
index of the diffuse gamma-ray emission.

Model predictions attempting to explain the excess flux in the
GeV region by assuming an
increased inverse Compton component (Porter \& Protheroe 1997, Pohl
\& Esposito 1998),
or by taking the contribution from unresolved SNRs into account
(Berezhko \& V\"olk, 2000), are given in the literature 
for different ranges in Galactic latitude $|b|$, and cannot be directly 
compared without assuming a latitude dependence of the
diffuse flux. In Fig. \ref{fig_results} the solid line gives the 
model prediction of Berezhko \& V\"olk 2000 scaled by a factor of 3, taking
into account the different latitude and longitude ranges used for the 
upper limit and the model and using the EGRET measurements at 20 GeV
as a guideline. 

Even for pessimistic assumptions of a rather
wide latitude dependence the model of Pohl \& Esposito (1998), if the inverse Compton
flux is
extrapolated from the 50 GeV range discussed in the paper to the TeV range,
exceeds the HEGRA limit. At TeV energies, one is still sufficiently
far from the Klein-Nishina regime such that the power-law extrapolation
should be valid (Porter \& Protheroe 1997). Of course, a break in the electron injection
spectrum could cause a corresponding break in the gamma spectrum between
50 GeV and 1 TeV, and could be used to make the model consistent 
with the experimental limit. 
\section{Concluding remarks}
A survey of the region $-10^\circ < b < 5^\circ$, 
$37^\circ < l < 43^\circ$ near the Galactic plane 
did not yield evidence for TeV gamma-ray point sources,
with typical flux limits of 10\% of the Crab flux.
In particular, 15 pulsars, 6 supernova remnants and one
unidentified EGRET source were not detected as strong TeV
sources.

A search for diffuse gamma-ray emission resulted in 
an upper limit of $6.1\cdot10^{-15}$ ph cm$^{-2}$ s$^{-1}$ sr$^{-1}$ MeV$^{-1}$ at 1 TeV, 
averaged over the region $38^\circ < l < 43^\circ, |b| < 2 ^\circ$ 
and assuming the spatial emission profile measured by the EGRET instrument. 
Since the analysis used to derive this limit
is only sensitive to the variation
of the diffuse flux with $b$, rather than its absolute value,
a distribution significantly wider than at EGRET energies
will increase the limit. Other variants of the data
analysis are sensitive to the absolute flux, but
give less stringent limits of $23.4$ and 
$10.4\cdot10^{-15}$ ph cm$^{-2}$ s$^{-1}$ sr$^{-1}$ MeV$^{-1}$.
The limit on the TeV gamma-ray flux can be used to 
derive a lower limit on the spectral index of the
diffuse radiation of 2.5, and to exclude models
which predict a strong enhancement of the diffuse 
flux compared to conventional mechanisms.
However the TeV flux limit is only about a factor of 1.5 larger than the
predicted flux from unresolved SNRs. A more sensitive survey should therefore
be able to test this prediction, together with a determination of the
"diffuse" TeV gamma-ray spectrum that is directly related to the Galactic
CR source spectrum.

\section*{Acknowledgments}

The support of the HEGRA experiment by the German Ministry for Research
and Technology BMBF and by the Spanish Research Council
CYCIT is acknowledged. We are grateful to the Instituto
de Astrof\'\i sica de Canarias for the use of the site and
for providing excellent working conditions. We gratefully
acknowledge the technical support staff of Heidelberg,
Kiel, Munich, and Yerevan. GPR acknowledges
receipt of a Humboldt Foundation postdoctoral fellowship.

\end{document}